\documentclass[12pt, draftclsnofoot, onecolumn]{IEEEtran}
\usepackage{cite}
\usepackage{float}
\usepackage{mathrsfs}
\usepackage{color}
\makeatletter

\newcommand{\Rmnum}[1]{\expandafter\@slowromancap\romannumeral #1@}

\makeatother
\usepackage{multirow}
\usepackage{algorithm}
\usepackage{algpseudocode}

\usepackage{graphicx}
\ifCLASSINFOpdf

\else

\fi

\usepackage[cmex10]{amsmath}
\usepackage{bm}

\usepackage{epstopdf}

\usepackage{cite}

\usepackage[cmex10]{amsmath}
\usepackage{array}

\usepackage{psfrag}
\usepackage{subfigure}
\usepackage{epsfig}
\usepackage{amssymb}

\usepackage{url}

\usepackage{booktabs} 

\begin{document}
%
%
\title{Massively Distributed Antenna Systems with Non-Ideal Optical Fiber Front-hauls: A Promising Technology for 6G Wireless Communication Systems}

\author{Lisu Yu, \IEEEmembership{Member, IEEE,}
        Jingxian Wu, \IEEEmembership{Senior Member, IEEE,}
        Andong Zhou, \\
        Erik G. Larsson, \IEEEmembership{Fellow, IEEE,}
        and Pingzhi Fan, \IEEEmembership{Fellow, IEEE}
\thanks{L. Yu is with the School of Information Engineering, Nanchang University, Nanchang 330031, P. R. China (e-mail: lisuyu@ncu.edu.cn).}
\thanks{J. Wu and A. Zhou are with the Department of Electrical Engineering, University of Arkansas, Fayetteville, AR 72701 USA (e-mails: wuj@uark.edu, az008@email.uark.edu).}
\thanks{E. G. Larsson is with the Department of Electrical Engineering, Link\"{o}ping University, 581 83 Link\"{o}ping, Sweden (e-mail: erik.g.larsson@liu.se).}
\thanks{P. Fan is with the Institute of Mobile Communications, Southwest Jiaotong University, Chengdu 611756, P. R. China (e-mail: pzfan@home.swjtu.edu.cn).}       
}

\maketitle

\begin{abstract}
Employing massively distributed antennas brings radio access points (RAPs) closer to users, thus enables aggressive spectrum reuse that can bridge gaps between the scarce spectrum resource and extremely high connection densities in future wireless systems. Examples include cloud radio access network (C-RAN), ultra-dense network (UDN), and cell-free massive multiple-input multiple-output (MIMO) systems. These systems are usually designed in the form of fiber-wireless communications (FWC), where distributed antennas or RAPs are connected to a central unit (CU) through optical front-hauls. A large number of densely deployed antennas or RAPs requires an extensive infrastructure of optical front-hauls. Consequently, the cost, complexity, and power consumption of the network of optical front-hauls may dominate the performance of the entire system. This article provides an overview and outlook on the architecture, modeling, design, and performance of massively distributed antenna systems with non-ideal optical front-hauls. Complex interactions between optical front-hauls and wireless access links require optimum designs across the optical and wireless domains by jointly exploiting their unique characteristics. It is demonstrated that systems with analog radio-frequency-over-fiber (RFoF) links outperform their baseband-over-fiber (BBoF) or intermediate-frequency-over-fiber (IFoF) counterparts for systems with shorter fiber length and more RAPs, which are all desired properties for future wireless communication systems.
\end{abstract}


\IEEEpeerreviewmaketitle

\section{Introduction}

\IEEEPARstart{W}{ith} the explosive growth of ubiquitous mobile services, the next generation wireless communication systems, such as beyond fifth generation (B5G) and sixth generation (6G) systems, will face formidable challenges imposed by the need for a large number of concurrent services with extremely high connection density. It is estimated that there are more than 9 billion connected devices at the end of 2018, and this number is predicted to reach 21 billion by 2020. With the rapid growth of connected devices, the connection density supported by a wireless system is expected to grow from 2,000 devices per km$^2$ to 1 million devices per km$^2$.

Massively distributed antenna system (DAS) is one of the most promising technologies to tackle the challenges imposed by high user density. A massive number of spatially distributed antennas or radio access points (RAPs) are deployed to provide geographical uniform services to users. DASs can be implemented in various architectures, such as cloud radio access network (C-RAN) \cite{wu2012green}, ultra-dense network (UDN) \cite{kamel2016ultra}, cell-free massive multiple-input multiple-output (MIMO) systems \cite{ngo2017cell}, etc. In all these architectures, we can move baseband unit (BBU) into a central unit (CU) to reduce both the network capital expenditure (CAPEX) and operating expense (OPEX).
The paradigm shift from cell-centric to user-centric can significantly improve the area spectral efficiency (ASE), thus effectively tackle the challenge imposed by the ever increasing connection density.

Either wireless or optical front-hauls can be used to connect the spatially distributed RAPs or antennas to CU. Wireless front-hauls are easy to setup, but are subject to harsh wireless propagation environment, especially in dense urban area where massively distributed antennas is needed the most. If the cell size is much smaller than the RAP-CU distance, then the communication bottleneck could shift from the wireless access links between RAP and user to the wireless front-hauls. In addition, the mutual interference between wireless access links and wireless front-hauls will further affect the performance of the network.

Optical front-hauls, on the other hand, have extremely high spectral efficiency, reliability, and security. Employing optical front-hauls requires an optical infrastructure, the construction of which are usually expensive and time consuming. However, once the infrastructure is setup, it is there to stay and can be used to serve many different applications for decades to come. In addition, optical front-hauls can be constructed by taking advantage of existing optical infrastructure pre-wired in buildings and industrial plants. Thus optical front-haul is a good long-term investment, and is a better front-haul technology for future wireless systems.

The combination of optical front-hauls and wireless access links forms the so-called fiber-wireless communications (FWC) \cite{lim2010fiber}, which enables us to push the RAPs as close to the user as possible. Due to the short distance between RAP and users, the wireless access links make up only a very small portion of the entire system. The power consumption, cost, and complexity of the optical front-hauls could easily dominate their wireless counterparts. It is thus critical to quantitatively identify the impacts of optical front-hauls over the entire network, thus achieve optimum designs of such networks by considering the complex interactions between the wireless and optical domains.

This article provides an overview and outlook on the architecture, modeling, design, and performance of massively distributed antenna systems with non-ideal optical front-hauls. The system architecture is modeled and analyzed by considering the interactions between optical front-hauls and wireless access links. Based on the modeling and analysis results, it is recognized that the optimum designs of such systems need to be performed by crossing the optical and wireless domains to jointly exploit the unique features and limits in both domains. Numerical results indicate that analog front-hauls with radio-frequency-over-fiber (RFoF) is more suitable for massively distributed antenna systems with a large number of antennas and relatively short length of optical fibers.

The remainder of this article is organized as follows. Section II provides an overview of the network architecture and operation scenarios of massively distributed antenna systems with optical front-haul. Detailed modeling and architecture of optical front-hauls are presented in Section III. Section IV discusses joint designs across optical and wireless domains that can benefit from the interactions between the two domains. Section V concludes the article.

\section{Network Architecture}

\subsection{Operation Scenarios}

\begin{figure}[!ht]
\begin{center}
\includegraphics[width=0.8\textwidth]{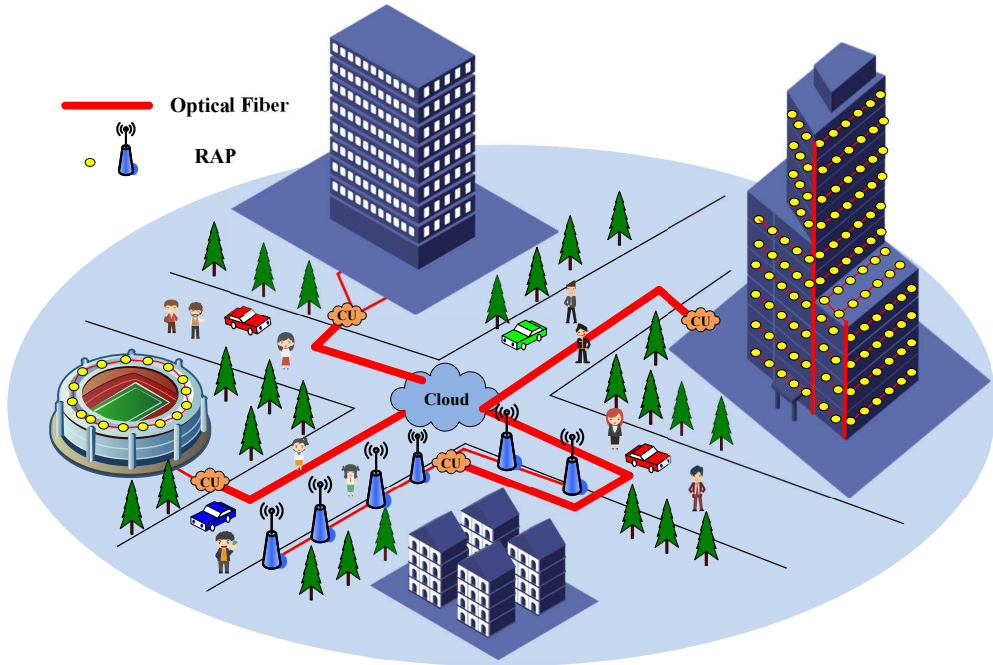}
\end{center}
\caption{System architecture of FWC systems with massively distributed antennas.}\label{fig:scenario}
\end{figure}

As shown in Fig. \ref{fig:scenario}, the typical operation scenarios of systems with massively distributed antennas include dense urban area such as downtown of big cities, sport venues such as football stadiums, or industrial environment such as big manufacturing plants. All these scenarios need to support massive number of concurrent connected devices with high density, which can benefit from the deployment of geographically distributed antennas.

The network can be constructed and deployed by taking advantage of existing infrastructure. The optical front-hauls can be implemented by utilizing existing optical fiber infrastructure. Many new commercial buildings and industrial plants already have optical fibers pre-wired during construction, or have plans for upgrading to optical fiber networks in the near future. It is estimated that about 54.8\% of medium to large commercial buildings in the United States have fiber-optic connectivity in 2017, and this number is increasing at a fast pace. The existing optical fiber infrastructure can be easily configured to support optical front-hauls of future wireless systems.

For example,  in a dense urban area, huge number of distributed antennas or RAPs can be installed on the windows, external walls, and roofs of high rise buildings by using the existing optical network in the building as optical front-hauls. Such configurations can convert external walls of buildings into a ``smart surface" as shown in Fig. \ref{fig:scenario}, which can provide reliable and geographically uniform services to users. Similarly, antennas or RAPs can be installed circling around the stands of a stadium, or along the shelves or cable conduits in an industrial plant.

Deploying a massive number of spatially distributed RAPs brings the radio signal sources much closer to the users. Such an approach is especially beneficial for millimeter wave (mm-wave) communications, where the signal propagation range is small but highly directional. Bringing RAPs closer to the users can improve communication quality-of-service (QoS) from the perspective of both coverage and interference management.

\subsection{C-RAN, UDN, Cell-free Massive MIMO?}

Systems with distributed antennas or distributed RAPs can be implemented in various forms, such as C-RAN, UDN, or cell-free massive MIMO. All these techniques rely on advanced front-haul techniques to achieve coordination and cooperation across geographically distributed antennas or RAPs, such that all users in the network can enjoy uniformly good services regardless of their actual locations. There are also noticeable differences among these techniques. Below is a brief summary of the main features and differences among C-RAN, UDN, and cell-free MIMO.

\begin{enumerate}
\item Architecture. In C-RAN and UDN, a few adjacent remote radio heads (RRHs) or RAPs cooperate to provide coverage over a small area via joint user association and beamforming. There is in general no cooperation among RRHs and RAPs serving different users. For cell-free massive MIMO, the entire distributed antenna array performs phase-coherent beamforming without explicit user association.

\item Interference. The performance of C-RAN and UDN are interference limited because of the interference from non-cooperating RRHs or RAPs. The interference level in cell-free massive MIMO is much lower due to the phase-coherent beamforming across the entire region.

\item Active user density. In C-RAN, each RRH is equipped with multiple antennas, and they can serve more active users than the number of RRHs. In UDN and cell-free massive MIMO, the number of RAPs or distributed antennas is in general larger than the number of active users.

\item RAP complexity. RRH in C-RAN in general has the highest complexity with multiple antennas. Cell-free massive MIMO can be implemented by using low complexity antennas.
\end{enumerate}

\section{Architecture of Optical Front-hauls}

The optical front-hauls can be classified into three categories, baseband-over-fiber (BBoF) \cite{thomas2013baseband}, intermediate-frequency-over-fiber (IFoF) \cite{ishimura20181}, and radio-frequency-over-fiber (RFoF) \cite{novak2016radio}. In an IFoF or RFoF scheme, analog signals are transmitted in the optical fiber, and they will suffer from distortions and attenuations introduced by the optical channels. The distortions are accumulated through the fiber links and propagated to the wireless links with analog signal transmission. On the other hand, optical fiber has negligible impact on a BBoF scheme with digital baseband signals transmitted in the fiber. However, the complexity and power consumption of RAPs in a BBoF or IFoF system are much higher than their RFoF counterparts due to complex signal processing and radio frequency (RF) up-conversion \cite{thomas2013baseband} performed at the RAP. Given the huge number of antennas or RAPs required to support high connection density, it is desirable to install antennas or RAPs with low cost, low complexity, and low power consumption.

\begin{figure}[!ht]
\begin{center}
\includegraphics[width=0.8\textwidth]{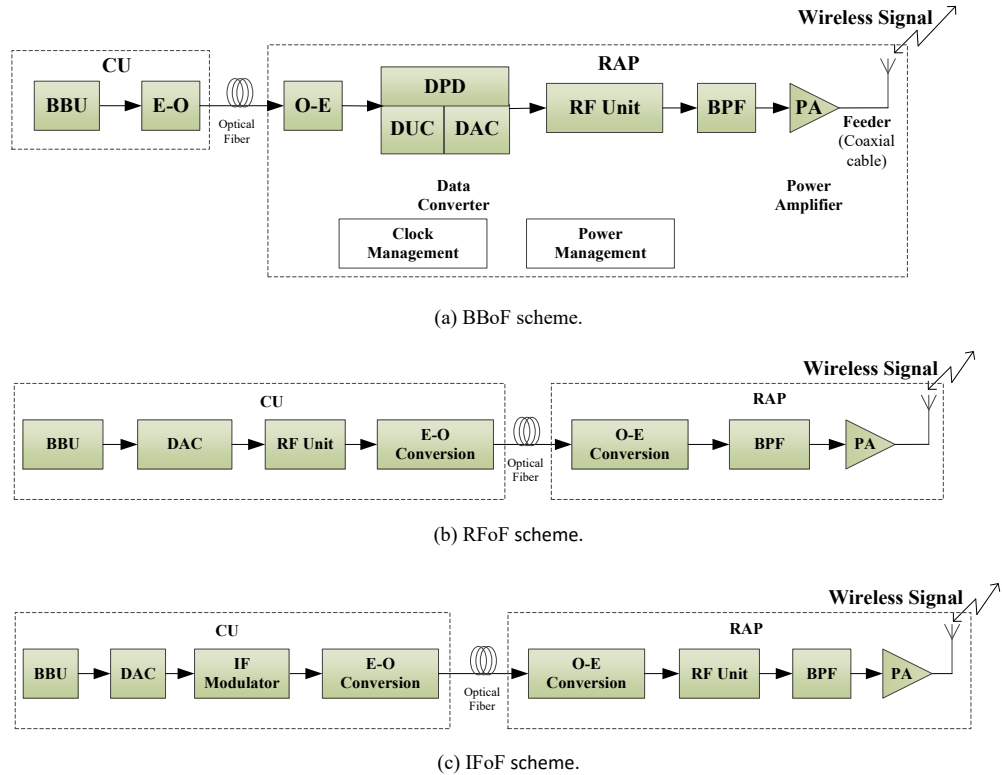}
\end{center}
\caption{Three kinds of FWC schemes architecture.} \label{FWC}
\end{figure}

\subsection{BBoF Transmission Scheme}

The architecture of a BBoF scheme through the CU to an RAP is shown in Fig. \ref{FWC}(a). It consists of the following modules \cite{yu2019energy}:

\begin{itemize}
\item  \textbf{Baseband Unit (BBU)} performs baseband processing with digital signal processors and microprocessors, such as coding, decoding, digital modulation, digital beamforming, channel estimation, equalization, and media access control (MAC), etc. 
\item \textbf{Electrical-Optical (E-O) and Optical-Electrical (O-E) interface} contains laser diodes and photo detectors to perform signal conversions between electrical and optical domains.
\item \textbf{Data converter unit (DCU)} contains some basic data processors, including the digital up converter (DUC), digital pre-distorter (DPD), and digital-to-analog converter (DAC).
\item \textbf{RF unit (RFU)} upconverts baseband signals to RF signals with local mixers.
\item \textbf{Bandpass filter (BPF)} is needed to limit the bandwidth of the RF signal to fit the allocated RF channel before transmission.
\item \textbf{Power amplifier (PA)} is used for RF signal amplification. The power consumption of PA depends on the transmission power and amplifier efficiency (AE). AE can be increased from 15\% to 25\% by digital pre-distortion. In analogy RFoF systems, there is no digital pre-distortion , and thus the PA efficiency is lower (about 15\%). In addition, a typical low-noise amplifier (LNA) has a low noise figure (NF) with 1 dB-3 dB, and it can provide a power gain, ${G_{{\text{PA}}}}$, of 10 (10 dB). 
\item \textbf{Feeder} normally uses a copper coaxial cable to connect the PA with the antenna, resulting in a loss of 3 dB.
\item \textbf{Clock management} provides synchronization function for each unit with synchronized clocks.
\item \textbf{Power supply and battery backup} provide uninterrupted power to the system. The loss of the two components is typically 10\%-15\% of the functional power consumption with different power technology.
\item \textbf{Cooling} is needed to keep the normal operation of the systems, and the cooling loss is typically from 0 to 40\% of the
functional power consumption, determined by the environmental conditions.
\end{itemize}

\subsection{RFoF Transmission Scheme}

Fig. \ref{FWC}(b) shows the architecture of an RFoF link. The RAP structure in an RFoF link is much simpler than its BBoF counterpart, because the baseband units and RF upconverter are located at the CU. Such an architecture can significantly reduce the cost and power consumption of the entire network with massively distributed antennas.

At the CU, BBU processes the digital information, the output digital signal is further processed and converted into analogy signal by using the data converter. The baseband signal is then upconverted into analogy RF signal through the RF unit. After the upconversion stage, optical modulation is applied to convert the RF signals into optical signals \cite{thomas2015performance}.
The optical signals are transmitted to the RAP by using low attenuation optical fibers with a loss at 0.3 dB/km. At the RAP, the optical signals are converted to RF signals with the O-E interface and BPF \cite{thomas2015performance}. 
It is noted that different from the BBoF scheme, there is no digital pre-distortion function performed in the RFoF or IFoF schemes, and the PA efficiency is usually around 15\%.

RF signals in the optical fiber suffers from optical propagation loss, chromatic dispersion, and optical noises. Since analog signal are transmitted in the system, these distortions are accumulated and propagated to the user equipment (UE) through the wireless links. Thus it is necessary to perform design of UDN with RFoF across both the optical and wireless domains.

\subsection{IFoF Transmission Scheme}

The architecture of an IFoF scheme is shown in Fig. \ref{FWC}(c). Different from the RFoF scheme above, the RF unit in IFoF is moved to the RAP at the IFoF link, and the IF modulator (IFM) is added into the CU. In IFoF schemes signals are first modulated to a lower intermediate frequency of a few hundred MHz.  The employment of intermediate frequency can relax the hardware requirements of high-speed optoelectronic devices.

In an IFoF scheme, we transmit the wireless signals at a low frequency of a few hundred MHz over the optical link. The use of intermediate frequency can help to relax the hardware requirements of high-speed optoelectronic. In addition, the impact of chromatic dispersion is relatively smaller to IF signals compared to RF signals. However, all of these advantages are achieved at the cost of a more complex RAP design, which needs to perform frequency upconversion with a stable local oscillator (LO) and mixers.

In summary, there are four main distinctions among the BBoF, IFoF and RFoF schemes.

\begin{itemize}
	\item {\bf RAP complexity.} The RAP in RFoF links is much simpler compared to that in BBoF and IFoF links. Consequently, the cost and power consumption of RAPs in an RFoF scheme is much lower than their BBoF and IFoF counterparts. This is especially desirable for systems with a massive number of geographically distributed RAPs. 

	\item {\bf RAP synchronization.} Synchronization among the spatially distributed RAPs can be easily achieved in RFoF systems through optical delays. The CU in RFoF systems can achieve precise synchronization among all RF signals by introducing different optical phase shifts based on the signal propagation delays before transmission. On the other hand, in BBoF or IFoF scheme, all RAPs must subscribe to the same clock in order to achieve synchronization of the RF signals. 

	\item {\bf Optical distortions.} Optical distortions and attenuations have significant impact on the design and performance of RFoF or IFoF schemes with analog signals transmitted in the optical fiber, yet they have negligible impact on BBoF systems with digital optical signals. In addition, RFoF signal suffers from severe chromatic dispersion.

	\item {\bf Noise accumulation.} The noise power in RFoF or IFoF scheme is in general larger than that in BBoF scheme due to the noise accumulated from CU to RAP. 
\end{itemize}

\section{Performance Comparisons of Optical Front-Hauls}

The performances of various optical front-haul techniques are compared in this section in terms of power consumption and total throughput of the overall system.

\subsection{Power Consumption Analysis}

In BBoF, the total power consumption can be calculated by adding up the power consumption of all modules in the block diagram given in Fig. \ref{FWC}(a). The typical values for commercially available devices are listed in Table \ref{table:Parameters} \cite{yu2019energy}.

\begin{table}[htbp]
\centering
\caption{Power Related Parameters in FWC Systems}\label{table:Parameters}
\begin{tabular}{|c|c|}  
  \hline
  \textbf{Parameters} &\textbf{Values}  \\
   \hline
  Power Consumption of BBU (${P_{{\text{BBU}}}}$) & 58 W \\
   \hline
  Power Consumption of IFM (${P_{\text{IFM}}}$) & 2 W \\
   \hline
  Power Consumption of DUC (${P_{{\text{DUC}}}}$) & 3 W \\
   \hline
  Power Consumption of DPD (${P_{{\text{DPD}}}}$) &  5 W \\
  \hline
  Power consumption of DAC (${P_{{\text{DAC}}}}$) & 2 W \\
   \hline
  Power Consumption of RF Unit (${P_{{\text{RFU}}}}$)  & 2 W \\
   \hline
 Power Consumption of Clock Management (${P_{{\text{CM}}}}$) & 1 W \\
  \hline    
  \multirow{3}{*}{PA Efficiency (${\mu _{{\text{PA}}}}$)} & 25\% (BBoF) \\
                                     & 15\% (RFoF) \\
                                     & 15\% (IFoF) \\
  \hline   
  Feeder Loss (${L_{{\text{feeder}}}}$) & 0.5  \\
   \hline
  Power Supply Loss (${L_{{\text{PS}}}}$) & 0.15  \\
   \hline
  Cooling Efficiency (${\mu _{\text{C}}}$) & 0.2  \\
   \hline
\end{tabular}
\end{table}

In BBoF, the total power consumption can be calculated by adding up the power consumption of all modules in the block diagram given in Fig. \ref{FWC}(a).

For RFoF and IFoF systems, the analog signals in the optical fiber are subject to power losses due to both fiber attenuation and chromatic dispersion. Fiber attenuation is proportional to fiber length at about 0.3 dB/km. Chromatic dispersion introduces different phase shifts or delays to optical signals of different wavelengths or frequencies. As a result, optical signals of different wavelengths arrive at the destination at slightly different time, and this results in a dispersion of the signal in the time domain.

\begin{figure}[!ht]
\centering
\includegraphics[width=0.8\textwidth]{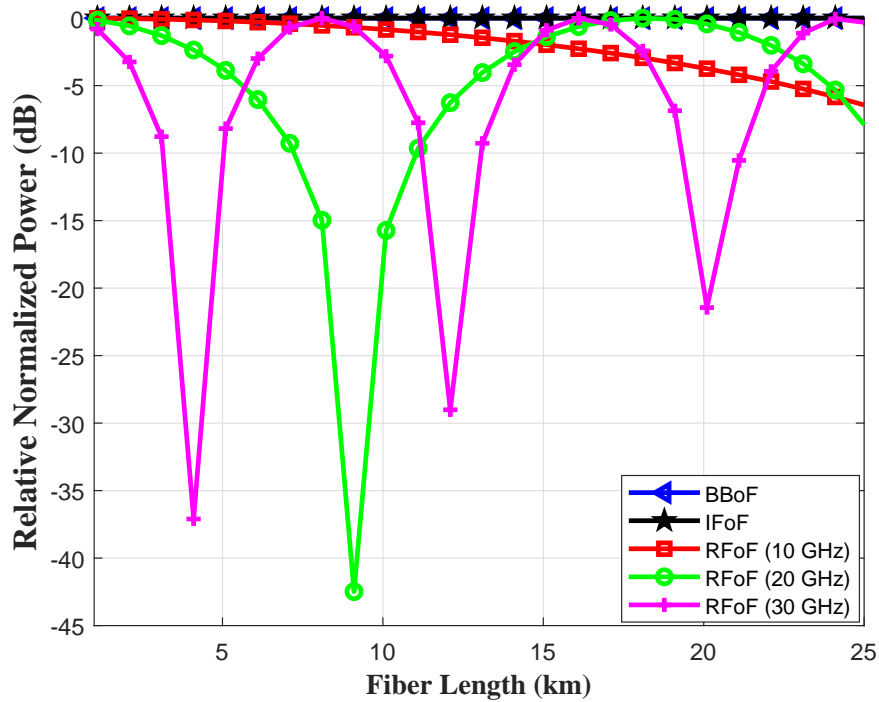} 
\caption{Normalized power variations comparisons among BBoF, IFoF and RFoF schemes versus the optical fiber length.}\label{fig:CD_comparion_distance}
\end{figure}

Fig. \ref{fig:CD_comparion_distance} shows the chromatic dispersion power loss for an RFoF system with intensity modulation (IM) and coherent homodyne detection \cite{yu2019energy}. In the figure, the wavelength of the optical carrier is 1553.6 nm, and the IF frequency is 125 MHz. RFoF suffers severe power losses due to chromatic dispersion, yet the impacts of chromatic dispersion on BBoF and IFoF systems are negligible. The power losses due to chromatic dispersion are more sensitive to fiber length at higher frequency. For example, increasing the fiber length from 1 to 4 km results in a power loss of 0.3 dB, 3 dB, and 37 dB for systems with RF carrier frequencies at 10 GHz, 20 GHz, and 30 GHz, respectively. On the other hand, for the 30 GHz system, there is negligible power losses when the fiber length is 8, 16, or 24 km. Therefore during the design of RFoF systems, the optical fiber lengths need to be carefully planned based on the RF carrier frequency to avoid significant power losses.

Several methods have been proposed to compensate power losses caused by chromatic dispersions \cite{xu2017analysis}. Many compensation methods rely on digital equalizations to reduce the negative impacts of chromatic dispersion, such as time-domain chromatic dispersion equalization (TDCDE), frequency-domain chromatic dispersion equalization (FDCDE), and least mean square (LMS) adaptive chromatic dispersion equalization (LACDE). Equalization-based compensation techniques generally have high complexity, and they may enhance the phase noise of the laser sources. Chromatic dispersion can also be compensated by employing dispersion-compensating fiber (DCF), the dispersion of which is negative in sign and much larger in magnitude compared to standard fiber. A short length  DCF can spliced into a longer length standard fiber to compensate dispersion of a standard single-mode fiber. In addition to the above techniques, we can also avoid significant power losses caused by chromatic dispersions by carefully planning on the length of the optical fibers based on the operating frequency, such that the nulls shown in Fig. \ref{fig:CD_comparion_distance} can be avoided.

\begin{figure}[!t]
\centering
\includegraphics[width=0.8\textwidth]{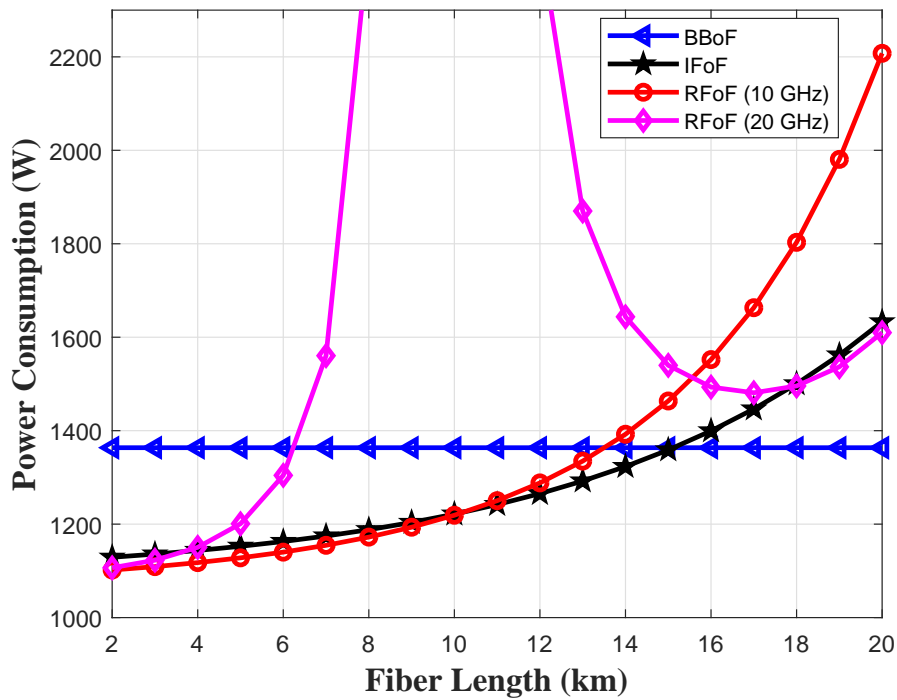}
\caption{Power consumption of different FWC schemes.}\label{fig:comparison_distance}
\end{figure}

Fig. \ref{fig:comparison_distance} compares the total power consumption of BBoF, IFoF and RFoF schemes. No chromatic distortion compensation is employed in the power evaluation. All three schemes have the same wireless transmission power, and the carrier bandwidth is set as 10 MHz. For the BBoF scheme, the total power consumption is independent of the optical fiber length. For the IFoF scheme, the total power consumption is an increasing function in fiber length mainly due to optical signal attenuation. The power consumption of RFoF scheme varies significantly with fiber length due to chromatic dispersion. The total power consumption of the RFoF schemes is less than that of the BBoF scheme when the fiber length is less than 13.5 km and 6 km, for systems with RF carrier frequencies 10 GHz and 20 GHz, respectively. Thus, an RFoF link is more suitable for short haul connection between CU and the RAP.

\subsection{Case Study}

We consider a simple case study to numerically demonstrate the impacts of the interactions between the optical and wireless domains in systems with massively distributed antennas. Two different scenarios are considered under the configurations of UDN and cell-free massive MIMO, respectively. For UDN, it is assumed that each RAP serves the closest UE. For cell-free massive MIMO, conjugate beamforming is performed across the entire antenna array \cite{ngo2017cell}.

\begin{figure}[!t]
\centering
\includegraphics[width=0.8\textwidth]{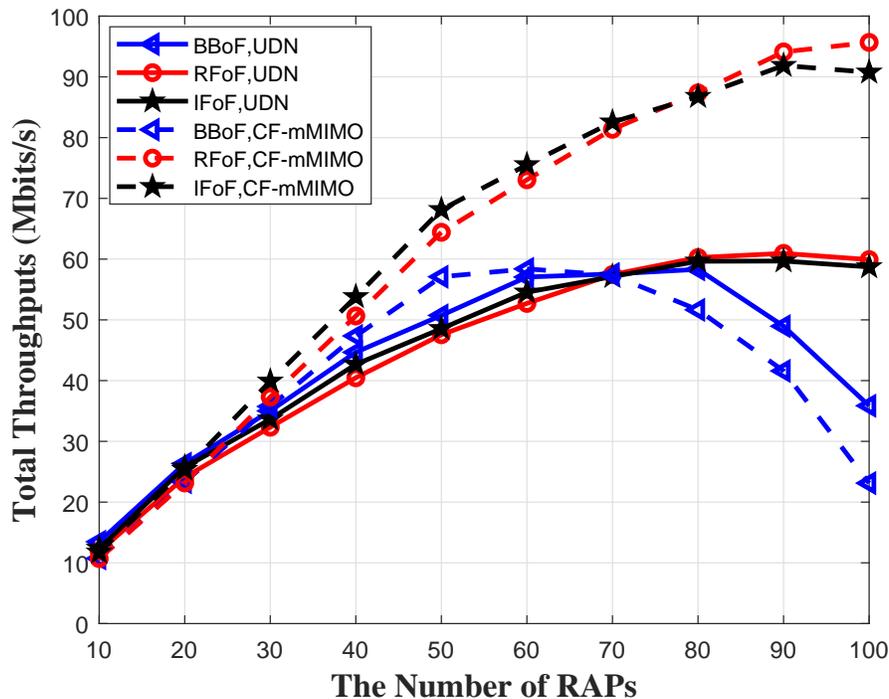}
\caption{Total throughput under different system configurations.}
\label{fig:throughput}
\end{figure}

Fig. \ref{fig:throughput} shows the throughput as a function of the number of RAPs $M$. In the simulations, we consider a large network with $M$ RAPs and $J$ UEs randomly distributed in an area of $1000{\text{m}} \times 1000{\text{m}}$ with $J=0.5M$. The upper bound of total power consumptions of all systems are the same at 2100 W. The CU is located 19 km away. 
The fiber chromatic dispersion coefficient is 17 ps/(nm$\cdot$km) with optical carrier wavelength 1553.6 nm. The data rate transmitted in optical fiber is 2.5 Gbit/s, the IF signal carrier frequency is 125 MHz, and the RF frequency is 20 GHz.

The throughput of all systems are concave in $M$ due to the increased overhead with more RAPs. The throughputs of RFoF and IFoF based systems are bigger than their corresponding BBoF counterparts, and the gap grows bigger at larger $M$. For the IFoF and RFoF schemes, the cell-free massive MIMO system achieves significant performance gains over their UDN counterparts. The performance gain is mainly due to the phase coherent conjugate beamforming used cell-free massive MIMO, which achieves significant reduction in interference across the entire network. Therefore, the IFoF and RFoF schemes achieve better performance for cell-free massive MIMO systems with a large number of RAPs.

\section{Joint Designs across Optical and Wireless Domains: Mixed Optical-Digital Beamforming}

The optimum designs of systems with massively distributed antennas must consider the complex interactions between the optical and wireless domains. Noise and distortions in IFoF or RFoF links are accumulated in the optical domain and propagated to the wireless domain, and this will have significant impacts on the design and performance of the overall system. As a result, the optical front-hauls must be designed by considering the characteristics of the wireless links, and vice versa.

One aspect that can directly benefit from cross-domain design is the development of beamforming, which is an essential component for all massively distributed antenna systems.
The spatially distributed RAPs form a distributed antenna array. Beamforming can thus be employed by the cooperating RAPs to steer and focus the signal to the desired UE. 

If the CU has perfect knowledge of the channel state information of the optical-wireless links of all RAPs, then digital beamforming can be performed to simultaneously increase signal power and reduce interference power. In an ideal setup, digital beamforming can be coherently performed across all RAPs to achieve the optimum performance by maximizing the signal-to-interference-plus-noise-ratio (SINR) at each UE \cite{li2004performance} at a given frequency. Various beamforming technqiues have been developed for C-RAN, UDN, and cell-free massive MIMO systems. In C-RAN and UDN, beamforming is usually performed together with user association, where each user is associated with one or more RAPs, and RAPs associated with different users in general transmit in a phase non-coherent manner without explicit cooperation. In cell-free massive MIMO, beamforming can be coherently performed across the entire antenna array such that all antennas form a phase-coherent antenna array. As a result, cell-free massive MIMO can reduce the overall interference floor due to the phase-coherent transmission. Most existing beamforming techniques are developed by assuming ideal optical front-hauls. However, non-ideal optical front-hauls may have significant impacts on beamforming development for systems with massively distributed antennas.

All above beamforming techniques require precise synchronization among all cooperating RAPs or antennas. In conventional distributed networks, the synchronization is usually achieved by requiring all RAPs subscribe to the same global clock, often from global position system (GPS) signals. The precision of clock signals from GPS signals might not meet the requirements of many digital beamforming schemes, which require signals from different RAPs arrive at the UE at the same symbol interval. The symbol interval could be on the order of nano-seconds for high throughput transmissions.
A small error in synchronization may cause corruption of the beam pattern, thus results in significant performance losses. In addition, the synchronization modules will increase the cost, complexity, and power consumption of RAPs.

We propose to perform mixed digital-optical beamforming to tackle the synchronization problem with a large number of distributed RAPs. Optical beamforming has been mainly used in radar systems or satellite communications with phased-array antennas. Optical beamforming enjoys the advantages of fast beam steering, large bandwidth, immunity to electromagnetic interference, high flexibility, low weight and size, and low cost \cite{vidal2012fast}. There are two main optical beamforming schemes, phase-only beamforming and true time-delay (TTD) beamforming \cite{frankel1998practical}. Phase-only beamforming is only applicable to narrow-band signal because of beam-squint, which is used to refer to the phenomenon that beam steering angles change with frequency. TTD beamforming, on the other hand, can be applied to wideband signals by achieving phase shifts that are linear in frequency. One possible implementation of TTD beamforming is using optical delay lines \cite{rideout2007true}, which introduces true time-delays to optical signals. However, existing optical beamforming techniques is usually designed by considering only the beam pattern of the signal component without considering interferences. Given the fact that the performance of massively distributed antenna system is usually dominated by the impacts of mutual interference among RAPs, beamforming should be able to simultaneously improve signal power and reduce interference power.

This can be achieved through mixed digital-optical beamforming, where we can combine the advantages of both digital and optical beamforming. In the baseband, digital beamformers can be designed to maximize the SINR. In the optical domain, TTD optical beamforming can then be designed to achieve two functions. First, TTD beamforming can be used to introduce linear phase to the RF signals to avoid beam-squint in wideband signals. Second, precise synchronization can be achieved by using optical delay lines at the CU to compensate different propagation delays, such that the cooperating signals can arrive at the desired UE at precisely the same time.

As an example, a mixed digital-optical beamforming scheme can be developed for cell-free massive MIMO systems by employing phase-coherent conjugate beamforming in the digital domain, and TTD optical beamforming in the optical domain. Unlike conventional conjugate beamforming that only considers the wireless channel, the digital beamforming coefficients in the cross-domain design should include the channel of both the optical and wireless channels. In addition, the incorporation of TTD optical beamforming can ensure that the phase-coherent signals from all antennas arrive at the desired user at precisely the same moment to ensure the phase coherence of signals from a large number of spatially distributed antennas. Furthermore, linear phase due to TTD beamforming can ensure the phase coherence is achieved over a wide spectrum range for wideband communications.

The mixed digital-optical beamforming approach optimizes system performance by taking advantage of the unique properties in both the optical and wireless domains, and it has the potential to achieve significant performance improvement over existing electrical beamforming schemes. We would like to point out that the proposed mixed digital-optical beamforming scheme is different from the hybrid digital-analog beamforming \cite{han2015large}, where analog phase-only beamformers are used in combination with digital beamforming to reduce the number of RF chains. As a result, the performance of joint digital-analog beamforming is always inferior to digital beamforming. On the other hand, mixed digital-optical beamforming could outperform digital beamforming, especially for wideband applications due to the linear phase with TTD beamforming in the optical domain.

\section{Conclusions}
Modeling and analysis of FWC systems with massively distributed RAPs have been studied in this paper. FWC systems with BBoF, IFoF, and RFoF front-hauls have been modeled and compared in terms of total power consumption and total throughput. The results highlighted the importance of cross-domain design that can jointly exploit the features and interactions across both optical and wireless domains. Mixed digital-optical beamforming has been proposed as candidate techniques for future wireless systems. Simulation results demonstrated that systems with RFoF front-hauls outperform their BBoF or IFoF counterparts for systems with shorter fiber length and more RAPs, which are all desired properties for future wireless communication systems.

\end{document}